\documentclass[12pt]{article}

\usepackage{graphics}
\usepackage{amssymb}
\usepackage{amsthm}

\newtheorem{theorem}{Theorem}
\newtheorem{lemma}{Lemma}

\newtheorem{example}{Example}
\newtheorem{remark}{Remark}

\usepackage{amsbsy}
\usepackage{epsfig}
\usepackage{caption}

\begin{document}

\title{Analytic Solutions for D-optimal Factorial Designs under Generalized Linear Models}

\author{Liping Tong\\ Loyola University Chicago
\and Hans W. Volkmer\\ University of Wisconsin-Milwaukee
\and Jie Yang\\ University of Illinois at Chicago}

\maketitle

\begin{abstract}
We develop two analytic approaches to solve D-optimal approximate designs under generalized linear models. The first approach provides analytic D-optimal allocations for generalized linear models with two factors, which include as a special case the $2^2$ main-effects model considered by Yang, Mandal and Majumdar (2012). The second approach leads to explicit solutions for a class of generalized linear models with more than two factors. With the aid of the analytic solutions, we provide a necessary and sufficient condition under which a D-optimal design with two quantitative factors could be constructed on the boundary points only. It bridges the gap between D-optimal factorial designs and D-optimal designs with continuous factors.
\end{abstract}


\section{Introduction}
\label{section_introduction}

Generalized linear models (McCullagh and Nelder, 1989; Dobson and Barnett, 2008) have been widely used for modeling responses coming from an exponential family including Binomial, Poisson, Gamma, and many other distributions. Under generalized linear models, a link function $g$ connects the expectation of the response $Y$ with a linear combination of factors, either qualitative or quantitative. For example, under a $k$-factor main-effects model,
$$g(E(Y)) = \beta_0 + \beta_1 x_1 + \cdots + \beta_k x_k\ ,$$
where $\beta_0, \beta_1, \ldots, \beta_k$ are regression coefficients, and $x_1, \ldots, x_k$ represent the levels of $k$ factors respectively. For many applications in agriculture, industry, clinical trials, etc, the experimenters are able to control the levels of factors in different runs of experiments to get more accurate estimates of $\beta_0, \beta_1, \ldots, \beta_k$~. Unlike the case of linear models, the information matrix in generalized linear models for the estimation of parameters usually depends on the unknown parameters (see Khuri, Mukherjee, Sinha and Ghosh (2006) for a good review). One solution solving the dependence is to use Chernoff (1953)'s local optimality approach in which the unknown parameters are replaced by assumed values.  Then different optimality criteria, such as $D$-, $A$-, $E$-, $c$-optimality, may be applied to the information matrix with assumed parameter values to obtain the corresponding optimal designs (see, for example, Stufken and Yang (2012)). Alternative solutions include Bayesian approach (Chaloner and Verdinelli, 1995), maximin criteria (Pronzato and Walter, 1988; Imhof, 2001), and sequential design (Ford, Titterington and Kitsos, 1989; Khuri, Mukherjee, Sinha and Ghosh, 2006).

One solution is to deal with quantitative or continuous factors. For typical applications, the factor level $x_j$ is restricted to the closed interval $[a_j, b_j], j=1, \ldots, k$. A design problem is to find a set $\{ ({\bf x}_i, p_i), i=1, \ldots, m\}$, where ${\bf x}_i=(x_{i1}, \ldots, x_{ik})'$, $i=1, \ldots, m$ are design points that are combinations of factor levels, and $p_i$'s are the proportions of experimental units assigned to the corresponding design points (see, for example, Atkinson, Donev and Tobias (2007) and Stufken and Yang (2012)). For the case of single quantitative factor, Sitter and Wu (1993) provided characterizations of $D$-, $A$- and $F$-optimal designs for binary response. Stufken and Yang (2012) showed that the locally optimal design could be constructed by solving an equation of a single variable. For the case of two or more quantitative factors, numerical algorithms are typically used for searching for locally optimal designs (Stufken and Yang, 2012; Woods, Lewis, Eccleston and Russell, 2006).

Another solution is to deal with qualitative factors or quantitative factors but with pre-specified finite number of design points. In this case, a design matrix $X$ that consists of $m$ design points is given and the design problem is to find the optimal allocation ${\bf p}=(p_1, \ldots, p_m)'$ assigned on the $m$ design points. Yang, Mandal and Majumdar (2012) considered locally D-optimal designs with binary response and two two-level factors. They provided analytic D-optimal allocations for some special cases only. Yang, Mandal and Majumdar (2013) considered locally D-optimal designs with binary response and $k$ two-level factors. They proposed a highly efficient numerical algorithm, {\it lift-one algorithm},  for searching locally D-optimal allocations. Yang and Mandal (2013) extended Yang, Mandal and Majumdar (2013)'s results for more general models and any pre-specified set of design points, which provided a potential tool to bridge the gap between qualitative factors and quantitative factors (see Section~\ref{section_continuous} for more details).

Although analytic solutions for optimal designs under generalized linear models are only available for some special cases, they are preferable to numerical solutions in terms of computation complexity and accuracy. For some applications (see Section~\ref{section_continuous} for an example), even highly efficient algorithms can not compete with an analytic solution. Among different criteria of optimal designs, D-optimality leads to maximization of a homogeneous polynomial for a large class of generalized linear models (Yang and Mandal, 2013), which is relatively easier to deal with. Following Yang and Mandal (2013), one aim of this paper is to develop analytic solutions for D-optimal design problems with pre-specified design matrix $X$.

This paper is organized as follows. In Section~\ref{section_22}, we utilize the variable elimination techniques in a system of polynomial equations to derive the analytic D-optimal allocation for the $2^2$ main-effects model, which answers the question left by Yang, Mandal and Majumdar (2012) and generalizes their results. In Section~\ref{section_twofactor}, we use the same techniques to find analytic D-optimal allocations for any four distinct design points of two factors. In Section~\ref{section_threefactor}, we develop another analytic approach to find D-optimal allocations with three or more factors. In Section~\ref{section_continuous}, we develop a necessary and sufficient condition under which only the four boundary points are needed for a D-optimal design with two continuous factors. With the aid of the analytic solutions developed in Section~\ref{section_22} and Section~\ref{section_twofactor}, we are able to interpret the condition in terms of the regression coefficients. In Section~\ref{section_simu}, we show by examples some advantages of analytic solutions over numerical answers.

\section{Analytic D-optimal Allocation under $2^2$ Main-effects Model}
\label{section_22}

Yang, Mandal and Majumdar (2012) considered a $2^2$ main-effects generalized linear model $g(E(Y))= \beta_0 + \beta_1 x_1 + \beta_2 x_2$ for binary response $Y$ with link function $g$ and design matrix
\begin{equation}\label{2^2maineffects}
X=\left(
\begin{array}{rrr}
1 &  1 &  1\\
1 &  1 & -1\\
1 & -1 &  1\\
1 & -1 & -1
\end{array}\right)
\end{equation}
which consists of four design points $(x_1, x_2) = (1, 1),\ (1, -1),\ (-1, 1)$, and $(-1, -1)$. A D-optimal approximate design (or allocation) is a 4-tuple $(p_1,$ $p_2,$ $p_3, p_4)$ that maximizes the determinant of the Fisher information matrix (up to a constant)
$$|X'WX|=16 (p_1 p_2 p_3 w_1 w_2 w_3 + p_1 p_2 p_4 w_1 w_2 w_4 + p_1 p_3 p_4 w_1 w_3 w_4 + p_2 p_3 p_4 w_2 w_3 w_4)$$
or equivalently that maximizes the objective function
$$f(p_1,p_2,p_3,p_4) = v_1p_2p_3p_4 + p_1 v_2 p_3 p_4 + p_1 p_2 v_3 p_4 + p_1 p_2 p_3 v_4\ ,$$
where $W={\rm Diag}\{p_1w_1, p_2w_2, p_3w_3, p_4w_4\}$, $p_i \geq 0$, $\sum_{i=1}^4 p_i =1$, $w_i > 0$, $v_i=1/w_i$, $i=1,2,3,4$. As pointed out by Yang and Mandal (2013), the D-optimal design obtained here is not just for binary response $Y$, but also for $Y$ that follows Poisson, Gamma, or other exponential family distributions with a single-parameter. Following their extended setup, $w_i = \nu(\eta_i)$, where $\eta_i = \beta_0 + \beta_1 x_{i1} + \beta_{i2}$ for the $i$th design point $(x_{i1}, x_{i2})$ and $\nu=\left((g^{-1})'\right)^2/r$ with $r(\eta) = r(\beta_0+\beta_1 x_1 + \beta_2 x_2)={\rm Var}(Y)$. Examples of $\nu$ include $\nu(\eta)=e^\eta/(1+e^\eta)^2$ for binary response and logit link, $\nu(\eta)=e^\eta$ for Poisson response and log link, etc. In order to find out locally D-optimal allocation $(p_1, p_2, p_3, p_4)$, $\beta_0, \beta_1, \beta_2$ are assumed to be known. Thus $w_i$ and $v_i$ are known positive constants for commonly used link functions.

In this section, we aim to solve the optimization problem
\begin{eqnarray}
& & \max f(p_1,p_2,p_3,p_4)\nonumber\\
\mbox{subject to} & & p_1 \geq 0, p_2 \geq 0, p_3 \geq 0, p_4\geq 0\label{22optimalproblem}\\
& & p_1 + p_2 + p_3 + p_4 = 1\nonumber
\end{eqnarray}
The solution always exists and is unique due to the strict log-concavity of $f$ (Yang, Mandal and Majumdar, 2012).

Without any loss of generality, we assume $0 < v_1 \leq v_2 \leq v_3 \leq v_4$. Yang, Mandal and Majumdar (2012, Theorem~1 \& Theorem~2) found analytic solutions for the following special cases:
\begin{itemize}
\item[(i)] If $v_4 \geq v_1+v_2+v_3$, then the solution is $p_1=p_2=p_3=1/3$, $p_4=0$.
\item[(ii)] If $v_4 < v_1+v_2+v_3$ and $v_1=v_2$, then the solution is
$$p_1 = p_2  = \frac{2v_1}{-2\delta_{12} + D_{12}},\>
p_3 = \frac{1}{2} + \frac{v_4 - v_3 - 4v_1}{2(-2\delta_{12} + D_{12})},\>
p_4 = \frac{1}{2} - \frac{v_4 - v_3 + 4v_1}{2(-2\delta_{12} + D_{12})},
$$
where $\delta_{12} = v_3 + v_4 - 4v_1$, $D_{12} = \sqrt{\delta_{12}^2 + 12 v_3 v_4}$~.
\item[(iii)] If $v_4 < v_1+v_2+v_3$ and $v_2=v_3$, then the solution is
$$p_1 = \frac{1}{2} + \frac{v_4 - v_1 - 4v_2}{2(-2\delta_{23} + D_{23})},\>
p_2 = p_3  = \frac{2v_2}{-2\delta_{23} + D_{23}},\>
p_4 = \frac{1}{2} - \frac{v_4 - v_1 + 4v_2}{2(-2\delta_{23} + D_{23})},$$
where $\delta_{23} = v_1 + v_4 - 4v_2$, $D_{23} = \sqrt{\delta_{23}^2 + 12 v_1 v_4}$~.
\item[(iv)] If $v_3=v_4$, then the solution is
$$p_1 = \frac{1}{2} + \frac{v_2 - v_1 - 4v_3}{2(-2\delta_{34} + D_{34})},\>
p_2 = \frac{1}{2} - \frac{v_2 - v_1 + 4v_3}{2(-2\delta_{34} + D_{34})},\>
p_3 = p_4  = \frac{2v_3}{-2\delta_{34} + D_{34}},$$
where $\delta_{34} = v_1 + v_2 - 4v_3$, $D_{34} = \sqrt{\delta_{34}^2 + 12 v_1 v_2}$~.
\end{itemize}

For the more common case $0 < v_1 < v_2 < v_3 < v_4 < v_1 + v_2 + v_3$, Yang, Mandal and Majumdar (2012) did not find an analytic solution. In this section, we derive an analytic solution for the last and most difficult case.

\begin{lemma}\label{2^2lemma}
Suppose $0 < v_1 \leq v_2 \leq v_3 \leq v_4 < v_1 + v_2 + v_3$. Then the solution $(p_1,p_2,p_3,p_4)$ maximizing (\ref{22optimalproblem}) satisfies $0<p_i<1$, $i=1,2,3,4$.
\end{lemma}

Lemma~\ref{2^2lemma} is actually a special case of Lemma~\ref{2klemmab} in Section~\ref{section_threefactor} whose proof is provided in Appendix. Based on Lemma~\ref{2^2lemma}, we obtain a necessary condition for the solution as a direct conclusion of the Karush-Kuhn-Tucker condition (Karush, 1939; Kuhn and Tucker, 1951).

\begin{lemma}\label{2^2conditionlemma}
Suppose $0 < v_1 < v_2 < v_3 < v_4 < v_1 + v_2 + v_3$. Then a necessary condition for $(p_1, p_2, p_3, p_4)$ to maximize (\ref{22optimalproblem}) is
\begin{equation}\label{22condition}
\frac{\partial f}{\partial p_1} = \frac{\partial f}{\partial p_2} = \frac{\partial f}{\partial p_3} = \frac{\partial f}{\partial p_4}\ .
\end{equation}
\end{lemma}

Note that the equations~(\ref{22condition}) are equivalent to $\partial f/\partial p_1 = \partial f/\partial p_4$, $\partial f/\partial p_2 = \partial f/\partial p_4$, and $\partial f/\partial p_3 = \partial f/\partial p_4$, that is,
\[
\left\{
\begin{array}{ccc}
(v_4 - v_1) p_2 p_3 + (v_3 p_2 + v_2 p_3) (p_4 - p_1) & = & 0\\
(v_4 - v_2) p_1 p_3 + (v_3 p_1 + v_1 p_3) (p_4 - p_2) & = & 0\\
(v_4 - v_3) p_1 p_2 + (v_2 p_1 + v_1 p_2) (p_4 - p_3) & = & 0
\end{array}
\right.
\]

According to Lemma~\ref{2^2lemma}, $p_i > 0$, $i=1,2,3,4$. Let $y_i = p_i/p_4>0$, $i=1,2,3$. Then $p_1 + p_2 + p_3 + p_4 =1$ implies $p_4 = 1/(y_1 + y_2 + y_3 + 1)$ and $p_i = y_i/(y_1+y_2+y_3+1), i=1,2,3$. Equations~(\ref{22condition}) are equivalent to
\begin{eqnarray}
(v_4 - v_1) y_2 y_3 + (v_3 y_2 + v_2 y_3) (1 - y_1) &=& 0\label{22con_a}\\
(v_4 - v_2) y_1 y_3 + (v_3 y_1 + v_1 y_3) (1 - y_2) &=& 0\label{22con_b}\\
(v_4 - v_3) y_1 y_2 + (v_2 y_1 + v_1 y_2) (1 - y_3) &=& 0 \label{22con_c}
\end{eqnarray}

After solving equation~(\ref{22con_c}) with respect to $y_3$, we get
\begin{equation}\label{y3}
y_3 = 1 + \frac{(v_4 - v_3) y_1 y_2}{v_2 y_1 + v_1 y_2},
\end{equation}
or equivalently $p_3 = p_4 + (v_4-v_3)p_1p_2/(v_2 p_1 + v_1 p_2)$. Then we substitute (\ref{y3}) for $y_3$ in equations~(\ref{22con_a}) and (\ref{22con_b}) and get
\begin{eqnarray}
v_2^2 y_1 (1 - y_1) + & &\nonumber\\
 v_2 [(v_1 + v_4 y_1)(1 - y_1) + (v_4 - v_1)y_1] y_2 + & &\nonumber\\
    \left[v_1 v_3 (1 - y_1) + (v_1 - v_3 y_1 + v_4 y_1) (v_4 - v_1)\right] y_2^2 &=& 0\label{22con_d}\\
v_2 y_1 [v_1 + (-v_2 + v_3 + v_4) y_1] + & &\nonumber\\
 \left[v_1^2 + 2 v_1 (-v_2 + v_4) y_1 + v_4 (-v_2 - v_3 + v_4) y_1^2\right] y_2 + & &\nonumber\\
   (- v_1)(v_1 + v_4 y_1) y_2^2 &=& 0 \label{22con_e}
\end{eqnarray}

After solving equation~(\ref{22con_e}) with respect to $y_2$, we get the only positive solution
\begin{equation}\label{y2}
y_2 = \frac{1}{2} + \frac{(v_3 - v_2)y_1}{2 (v_1 + v_4y_1)} - \frac{(v_2 + v_3 - v_4) y_1}{2 v_1} + \frac{\sqrt{D_2}}{2 v_1(v_1 + v_4 y_1)}
\end{equation}
where $D_2 = \left[(v_1 + v_4 y_1)^2 - (v_3 - v_2) v_4 y_1^2\right]^2 -  4 v_2 (v_4 -v_3) (v_1 + v_4 y_1)^2 y_1^2$~. We then replace $y_2$ with (\ref{y2}) in equation~(\ref{22con_d}) and simplify the expression into
\begin{equation}\label{22solution}
c_0 + c_1 y_1 + c_2 y_1^2 + c_3 y_1^3 + c_4 y_1^4 = 0
\end{equation}
where
\begin{eqnarray*}
c_0 &=& 2 v_1^3 (-v_1 + v_2 + v_3 + v_4)\>\>\> > \>\>\>0\\
c_1 &=&  v_1^2\left[(-v_1-v_2+v_3+v_4)^2 + 4 (v_4 - v_1) (v_2 + v_4)\right] \>\>\> > \>\>\>0\\
c_2 &=& 2 v_1 v_4 \left[2 (v_1-v_4)^2 - (v_2-v_3)^2 - (v_1+v_4)(v_2+v_3)\right]\\
c_3 &=& v_4^2 \left[(v_1-v_2+v_3-v_4)^2 - 4 (v_4 - v_1) (v_1 + v_2)\right]\\
c_4 &=& 2 (v_1 + v_2 + v_3 - v_4) v_4^3 \>\>\> > \>\>\>0
\end{eqnarray*}

\begin{lemma}\label{y1>1}
There is one and only one $y_1>1$ solving equation~(\ref{22solution}).
\end{lemma}

The proof of Lemma~\ref{y1>1} is given in Appendix. According to the solutions provided by the software {\tt Mathematica}, the largest root of equation~(\ref{22solution}) after simplification is
\begin{equation}\label{y1solution}
y_1=-\frac{a_3}{4}+\frac{\sqrt{A_1}}{2} + \frac{\sqrt{C_1}}{2}\ ,
\end{equation}
where $a_0=c_0/c_4$, $a_1=c_1/c_4$, $a_2=c_2/c_4$, $a_3=c_3/c_4$,
\begin{eqnarray*}
A_1 &=& -\frac{2  a_2}{3}+\frac{ a_3^2}{4}+\frac{G_1}{3\times 2^{1/3}}\ ,\\
C_1 &=& -\frac{4 a_2}{3}+\frac{a_3^2}{2}-\frac{G_1}{3\times 2^{1/3}} + \frac{-8 a_1+4 a_2 a_3-a_3^3}{4 \sqrt{A_1}}\ ,\\
G_1 &=& \left(F_1-\sqrt{F_1^2-4 E_1^3}\right)^{1/3}+ \left(F_1+\sqrt{F_1^2-4 E_1^3}\right)^{1/3}\ ,\\
E_1 &=& 12 a_0+a_2^2-3 a_1 a_3\ ,\\
F_1 &=& 27 a_1^2-72 a_0 a_2+2 a_2^3-9 a_1 a_2 a_3+27 a_0 a_3^2\ .
\end{eqnarray*}
Note that the calculation of $G_1$, $A_1$, $C_1$ and thus $y_1$ should be regarded as operations among complex numbers since the expression under square root could be negative. Nevertheless, $y_1$ at the end would be a real number. That is, all the imaginary parts will be canceled out. Now we are able to provide the analytic solution for the last case of the optimization problem~(\ref{22optimalproblem}).

\begin{theorem}\label{thm:22solution}
Consider the optimization problem~(\ref{22optimalproblem}).
\begin{itemize}
\item[(v)] If $0 < v_1 < v_2 < v_3 < v_4 < v_1 + v_2 + v_3$, then the unique solution can be calculated analytically as follows
\begin{itemize}
\item[(1)] calculate $y_1>1$ according to formula~(\ref{y1solution});
\item[(2)] calculate $y_2>1$ according to formula~(\ref{y2});
\item[(3)] calculate $y_3>1$ according to formula~(\ref{y3});
\item[(4)] $p_i = \frac{y_i}{y_1+y_2+y_3+1}$, $i=1,2,3$; $p_4 = \frac{1}{y_1 + y_2 + y_3 + 1}$~.
\end{itemize}
\end{itemize}
\end{theorem}

\section{General Case of Two Factors}
\label{section_twofactor}

In this section, we consider a more general setup of the two factors $x_1$ and $x_2$. The design points are not restricted to $(1, 1),\ (1, -1),\ (-1, 1),\ (-1, -1)$ any more. Suppose there are four distinct design points under consideration. The design matrix $X$ in this section could be written as
\begin{equation}\label{Xgeneral}
X=\left(
\begin{array}{rrr}
1 &  x_{11} &  x_{21}\\
1 &  x_{12} &  x_{22}\\
1 &  x_{13} &  x_{23}\\
1 &  x_{14} &  x_{24}
\end{array}\right).
\end{equation}
According to Yang, Mandal and Majumdar (2013, Lemma~3.1), the objective function of a D-optimal design is
$$|X'WX|=w_1w_2w_3w_4\left(p_1 p_2 p_3 u_4 + p_1 p_2 p_4 u_3 + p_1 p_3 p_4 u_2 + p_2 p_3 p_4 u_1\right)\ ,$$
where $u_4=|X[1,2,3]|^2/w_4$, $u_3=|X[1,2,4]|^2/w_3$, $u_2=|X[1,3,4]|^2/w_2$, $u_1=|X[2,3,4]|^2/w_1$, and $X[i_1,i_2,i_3]$ represents the $3\times 3$ submatrix consisting of the $i_1$th, $i_2$th, $i_3$th rows of $X$.

The design problem is to maximize $|X'WX|$, which is equivalent to maximizing the objective function
$$f_u(p_1, p_2, p_3, p_4) = u_1p_2p_3p_4 + p_1 u_2 p_3 p_4 + p_1 p_2 u_3 p_4 + p_1 p_2 p_3 u_4\ .$$
The only difference between $f_u$ and $f$ in Section~\ref{section_22} is that $u_1,u_2,u_3,u_4$ could be $0$. Since the rows of $X$ are required to be distinct, then ${\rm rank}(X)\geq 2$. We provide the analytic D-optimal allocation ${\mathbf p}=(p_1, p_2, p_3, p_4)'$ which maximizes $|X'WX|$ or $f_u$ in three cases as follows.

\bigskip
{\it Case 1:} ${\rm rank}(X)=2$. In this case, one column of $X$ can be written as a linear combination of the other two columns. The model essentially has only one factor. It's a degenerated case such that $|X'WX|\equiv 0$. Mathematically, any allocation $(p_1, p_2, p_3, p_4)'$ is a solution maximizing $|X'WX|$.

\bigskip
{\it Case 2:} ${\rm rank}(X)=3$ and one row of $X$ can be written as a linear combination of two other rows. It can be verified that there is one and only one $u_i=0$ in this case. For example, if $\alpha_4 = a\alpha_2+b\alpha_3$, where $\alpha_i$ represents the $i$th row of $X$, then $u_1=0$ while $u_2 >0, u_3 >0, u_4>0$.
Without any loss of generality, assume $0=u_1 < u_2\leq u_3 \leq u_4$. Using the same analytic approach as in Section~\ref{section_22}, we get
\begin{itemize}
\item[{\it (2a)}] If $u_4 \geq u_2 + u_3$, then the solution is $p_1=p_2=p_3=1/3$, $p_4=0$.
\item[{\it (2b)}] If $u_4 < u_2 + u_3$ and $u_2=u_3$, then the solution is
$$p_1=\frac{1}{3},\>\>\>  p_2=p_3=\frac{2u_2}{3(4u_2-u_4)}, \>\>\> p_4=\frac{1}{2}-\frac{4u_2+u_4}{6(4u_2-u_4)}\ .$$
\item[{\it (2c)}] If $u_3=u_4$, then the solution is
$$p_1=\frac{1}{3},\>\>\> p_2=\frac{1}{2}-\frac{u_2+4u_3}{6(4u_3-u_2)},\>\>\> p_3=p_4=\frac{2u_3}{3(4u_3-u_2)}\ .$$
\item[{\it (2d)}] If $0=u_1 < u_2 < u_3 < u_4 < u_2 + u_3$, let $y_i = p_i/p_4 > 0$, $i=1,2,3$. Following the calculations in Section~\ref{section_22}, the equation parallel to (\ref{y3}) is $y_3 = 1 + (u_4-u_3)y_2/u_2$~. After solving the equation parallel to (\ref{22con_e}), we get
\begin{equation}\label{y2special}
y_2 = \frac{u_2(u_3+u_4-u_2)}{u_4(u_2+u_3-u_4)}\ .
\end{equation}
We then substitute (\ref{y2special}) for $y_2$ in an equation parallel to (\ref{22con_d}) and solve for $y_1$. The only positive solution is
$$y_1 = 1 + \frac{u_4^2 - (u_2 - u_3)^2}{2u_4(u_2 + u_3 - u_4)} = \frac{2 u_2 u_3 + 2 u_2 u_4 + 2 u_3 u_4 - u_2^2 - u_3^2 - u_4^2}{2u_4(u_2 + u_3 - u_4)}\ .$$
It can be verified that $y_1 > 1$. Then
$$y_3 = 1 + \frac{(u_3 + u_4 - u_2)(u_4-u_3)}{(u_2+u_3-u_4)u_4} = \frac{u_3(u_2 + u_4 - u_3)}{u_4(u_2 + u_3 - u_4)}\ .$$
Since $p_i = y_i/(y_1 + y_2 + y_3 + 1)$, $i=1,2,3$ and $p_4=1/(y_1 + y_2 + y_3 + 1)$, then the solutions is $p_1 = 1/3$, $p_2 = 2 u_2 (u_3 + u_4 - u_2)/(3\Delta)$, $p_3 = 2 u_3 (u_2 + u_4 - u_3)/(3\Delta)$, $p_4 = 2 u_4 (u_2 + u_3 - u_4)/(3\Delta)$, where $\Delta  = 2 u_2 u_3 + 2 u_2 u_4 + 2 u_3 u_4 - u_2^2 - u_3^2 - u_4^2 = (\sqrt{u_2}+\sqrt{u_3}+\sqrt{u_4}) (\sqrt{u_2}+\sqrt{u_3}-\sqrt{u_4}) (\sqrt{u_2}+\sqrt{u_4}-\sqrt{u_3})(\sqrt{u_3}+\sqrt{u_4}-\sqrt{u_2}) > 0$, since $0<u_2<u_3<u_4<u_2+u_3$~.
\end{itemize}

\begin{remark}
{\rm
If we go back to the formulas provided in cases (i)$\sim$(v) in Section~\ref{section_22} and let $v_1$ go to $0$, we can derive the same formulas listed in cases (2a), (2b), (2c) from cases (i), (iii), and (iv) respectively. However, if one wants to derive case~(2d) here from case~(v) in Section~\ref{section_22} directly, one will see the formula of $y_2$ in (\ref{y2special}) is not equal to the limit $(v_4-v_2)/v_4$ of (\ref{y2}) as $v_1$ goes to $0$. It is actually another example that the solution of a polynomial may not change continuously along with the changes of its coefficients.
}
\end{remark}

{\it Case 3:}  ${\rm rank}(X)=3$ and no row of $X$ can be written as a linear combination of two other rows. In this case, $u_i>0$, $i=1,2,3,4$. The solution could be obtained from Section~\ref{section_22} by replacing $u_i$ with $v_i$, $i=1,2,3,4$.

\begin{example}\label{boundaryexample}{\rm
Motivated by applications with two quantitative factors, where typically the two factors are bounded, e.g. $x_1 \in [a_1, b_1]$, $x_2 \in [a_2, b_2]$, a special case of the design matrix $X$ in (\ref{Xgeneral}) may consist of four boundary points, that is
\[
X=\left(
\begin{array}{rrr}
1 &  b_1  &  b_2\\
1 &  b_1  &  a_2\\
1 &  a_1  &  b_2\\
1 &  a_1  &  a_2
\end{array}\right).
\]
Then $|X[1, 2, 3]| = |X[1, 2, 4]| = - (b_1 - a_1)(b_2 - a_2)$, $|X[1, 3, 4]|$ $=$ $|X[2, 3, 4]|$ $=$ $(b_1 - a_1)(b_2 - a_2)$. In this case, the objective function of a D-optimal design is $|X'WX|=(b_1-a_1)^2(b_2-a_2)^2w_1w_2w_3w_4 (p_1 p_2 p_3 v_4 + p_1 p_2 p_4 v_3 + p_1 p_3 p_4 v_2$ $+$ $p_2 p_3 p_4 v_1)$, which is proportional to $p_1 p_2 p_3 v_4 + p_1 p_2 p_4 v_3 + p_1 p_3 p_4 v_2 + p_2 p_3 p_4 v_1$~. Therefore, the D-optimal design in this case takes exactly the same form as the solution in Section~\ref{section_22} in term of $v_1, v_2, v_3, v_4$, although the $v_i$'s here do depend on $a_1, b_1, a_2, b_2$.
}\end{example}

\section{Case with Three Factors or More}
\label{section_threefactor}

In this section, we consider design problems with more than two factors. For example, in the generalized linear model $g(E(Y)) = \beta_0 + \beta_1 x_1 + \beta_2 x_2 + \beta_3 x_3 + \beta_{12}x_1x_2 + \beta_{13}x_1 x_3 + \beta_{23} x_2x_3$, there are three factors and seven parameters. If $8$ distinct design points are pre-specified, the design matrix $X$ would be $8\times 7$ with rows in the form of $(1,x_1,x_2,x_3,x_1x_2,x_1x_3,x_2x_3)$.

In general, for a locally D-optimal design problem with a pre-specified $n\times d$ design matrix $X$, the determinant $|X'WX|$ is an order-$d$ homogeneous polynomial of $p_1, \ldots, p_n$ (see Lemma~3.1 of Yang and Mandal (2013)):
$$|X'WX| = \sum_{1\leq i_1 < \cdots < i_d \leq n} |X[i_1, \ldots, i_d]|^2 \cdot p_{i1} w_{i1}\cdots p_{id} w_{id}~,$$
where $X[i_1, \ldots, i_d]$ represents the $d\times d$ sub-matrix consists of the $i_1\mbox{th}, \ldots, i_d\mbox{th}$ rows of $X$. Numerical approaches were commonly used to search for the optimal allocation ${\mathbf p} = (p_1, \ldots, p_n)'$.

The analytic approach we developed in Section~\ref{section_22} is to eliminate variables in a system of polynomial equations (see, for example, Chapter~2 in Cox, Little and O'Shea (2005) for more results from algebraic geometry). Through that way, we may obtain a polynomial equation of one variable $p_1$. However, if the number of factors $m$ becomes large the degree of the polynomial will become large and its coefficients will be complicated polynomials of the variables $v_i$'s. It will be almost impossible to use the method in Section~\ref{section_22} for large $m$.

In this section, we provide another analytic approach for a class of design problems with two or more factors and a pre-specified design matrix. More specifically, we consider the D-optimal design problem with a pre-specified $n\times (n-1)$ design matrix $X$, that is, $X$ consists of $n$ distinct rows for $(n-1)$ parameters. We assume that $X$ is of full rank, that is, of rank $(n-1)$. Otherwise, one could reduce the number of parameters by model reparametrization. It should be noted that the design problem with a pre-specified $n\times n$ design matrix leads a trivial optimization problem since it always yields $p_1=p_2=\cdots=p_n=1/n$ as an optimal allocation.

To simplify the situation, we first assume that no row of $X$ can be written as a linear combination of $(n-2)$ other rows. In other words, any $(n-1)$ rows of $X$ are linearly independent, which implies that $|X[i_1, \ldots, i_{n-1}]|\neq 0$ for any $1\leq i_1 < \cdots < i_{n-1} \leq n$. This assumption will be removed later this section.

Under the assumptions above, the D-optimal allocation problem, that is, to find out the best ${\mathbf p} = (p_1, \ldots, p_n)'$ maximizing $|X'WX|$, is equivalent to the optimization problem
\begin{eqnarray}
& & \max f(p_1,p_2,\ldots, p_n)=p_1p_2\cdots p_n\sum_{j=1}^n \frac{v_j}{p_j}\nonumber\\
\mbox{subject to} & & p_i \geq 0,\ i=1,\ldots,n\label{2koptimalproblem}\\
& & p_1 + p_2 + \cdots + p_n = 1\nonumber
\end{eqnarray}
where $v_j = |X[1,\ldots, j-1,j+1,\ldots, n]|^2 w_1\cdots w_{j-1}w_{j+1}\cdots w_n > 0$, $j=1,\ldots,n$. Note that we denote $p_1p_2\cdots p_n\frac{v_j}{p_j}=p_1\cdots p_{j-1} v_j p_{j+1}\cdots p_n$ at $p_j=0$.

\begin{example}\label{2kfullfactorialexample}
{\rm
Suppose there are $k$ two-level ($-1$ or $+1$) factors. Let $X$ be the $2^k\times (2^k-1)$ matrix whose rows include all combinations of the $k$ factors and whose columns include the $k$ main effects and all interactions but the one of order $k$. Then $|X[1,\ldots, j-1,j+1,\ldots, n]|^2=2^{k(2^k-2)}$ for all $j$, where $n=2^k$. In other words, the D-optimal allocation design problem takes the form of (\ref{2koptimalproblem}) with $v_j = 2^{k(2^k-2)} w_1\cdots w_{j-1}w_{j+1}\cdots w_n > 0$, $j=1, \ldots, 2^k$. A special case is the $2^2$ main-effects model where $X$ is given by (\ref{2^2maineffects}).
}
\end{example}

Now we consider the optimization problem~(\ref{2koptimalproblem}) with $v_j > 0$, $j=1, \ldots, n$. Without any loss of generality, we assume $0 < v_1 \leq v_2 \leq \cdots \leq v_n$~. Based on a similar proof as the one for Theorem~1 in Yang, Mandal and Majumdar (2012), we obtain
\begin{lemma}\label{2klemma}
If $v_n \geq \sum_{j=1}^{n-1} v_j$, then $f(p_1, \ldots, p_n)$ attains its maximum $v_n/(n-1)^{n-1}$ only at $p_1=\cdots=p_{n-1}=\frac{1}{n-1}$, $p_n=0$.
\end{lemma}
Otherwise, if none of $v_i$ is greater than the sum of the others, we have the result below to guarantee the solution must be an interior point. Proofs for both lemmas can be found in Appendix.

\begin{lemma}\label{2klemmab}
If $v_n < \sum_{j=1}^{n-1} v_j$, then the $\mathbf{p}=(p_1, \ldots, p_n)'$ maximizing $f(\mathbf{p})$ must satisfy $p_i>0$ for all $i$.
\end{lemma}

Note that both Lemma~\ref{2klemma} and Lemma~\ref{2klemmab} are valid even if $0 = v_1 = \cdots = v_l < v_{l+1} \leq \cdots \leq v_n$ for some $1\leq l\leq n-3$, which are needed later this section.

Now we consider the case $0 < v_1 \leq v_2 \leq \cdots \leq v_n < \sum_{j=1}^{n-1} v_j$. Due to Lemma~\ref{2klemmab} and the Karush-Kuhn-Tucker condition, a necessary condition under which $\mathbf{p} = (p_1, \ldots, p_n)'$ maximizes $f$ is
\begin{equation}\label{2kcondition}
\frac{\partial f}{\partial p_1} = \cdots = \frac{\partial f}{\partial p_n} =\lambda
\end{equation}
for some constant $\lambda$. Since $\frac{\partial f}{\partial p_i} = \frac{p_1 p_2 \ldots p_n}{p_i}\left(\sum_{j=1}^n\frac{v_j}{p_j} - \frac{v_i}{p_i}\right)$, $i=1,\ldots,n$, the equations can be written in its matrix form
$$(J-I)\left(\frac{v_1}{p_1},\ldots, \frac{v_n}{p_n}\right)' = \frac{\lambda}{p_1\cdots p_n}(p_1, \ldots, p_n)',$$
where $J$ is the $n$ by $n$ matrix with all entries equal to $1$ and $I$ is the $n$ by $n$ identity matrix. Since $(J-I)^{-1} = \frac{1}{n-1}J - I$, we get the equivalent equations
\begin{equation}\label{2kmu0}
\frac{v_i}{p_i} = \frac{\lambda}{p_1\cdots p_n}\left(\frac{1}{n-1} - p_i\right), \>\>\> i=1,\ldots, n,
\end{equation}
or equivalently
\begin{equation}\label{2kmu}
\frac{p_i\left(\frac{1}{n-1} - p_i\right)}{v_i} = \frac{\mu}{4(n-1)^2},\>\>\> i=1,\ldots,n,
\end{equation}
where $\mu=4(n-1)^2p_1\cdots p_n/\lambda$ does not depend on $i$. It can be verified that
$\mu>0$ and $0<p_i<\frac{1}{n-1}$ for all $i$. Note that $f(\mathbf{p})/(v_1\cdots v_n)$ is a symmetric function of $p_1/v_1, \ldots, p_n/v_n$. Due to the assumption $0<v_1\leq v_2 \leq \cdots \leq v_n < v_1 + \cdots + v_{n-1}$, it follows that $p_1 \geq p_2 \geq \cdots \geq p_n > 0$.

For a given $\mu>0$, we solve the quadratic equations~(\ref{2kmu}) and get two possible solutions for $p_i$,
$$p_{i+}=\frac{1+\sqrt{1-\mu v_i}}{2(n-1)},\>\>\> p_{i-}=\frac{1-\sqrt{1-\mu v_i}}{2(n-1)},\>\>\> i=1,\ldots, n.$$
Note that $p_{1+} \geq p_{2+} \geq \cdots \geq p_{n+} \geq \frac{1}{2(n-1)} \geq p_{n-} \geq \cdots \geq p_{2-} \geq p_{1-}$~. Since $0 < p_i < \frac{1}{n-1}$ for all $i$, there is at most one $p_i$ that takes the value of $p_{i-}$ (otherwise $\sum_i p_i <1$). Therefore, either $p_i=p_{i+}$ for all $i$, or $p_i=p_{i+}$ for $i=1,\ldots, n-1$ but $p_n=p_{n-}$. Both cases are possible. For examples, let $n=4$, then $p_4=p_{4+}$ if $(v_1, v_2, v_3, v_4)=(5,5,6,7)$; $p_4=p_{4-}$ if $(v_1,v_2,v_3,v_4)$ $=$ $(1,1,2,3)$.

To find out $\mu$, we consider two functions as follows
\begin{eqnarray*}
h_1(\mu) &=& \sum_{j=1}^n \sqrt{1-\mu v_j}\\
h_2(\mu) &=& \sum_{j=1}^{n-1} \sqrt{1-\mu v_j} - \sqrt{1-\mu v_n}
\end{eqnarray*}
defined for $0\leq \mu \leq v_n^{-1}$. Note that $\sum_{i=1}^n p_{i+}=1$ implies $h_1(\mu)=n-2$;
$\sum_{i=1}^{n-1} p_{i+} + p_{n-}=1$ leads to $h_2(\mu)=n-2$.

\begin{theorem}\label{2ktheorem}
Assume that $0 < v_1 \leq v_2 \leq \cdots \leq v_n < \sum_{j=1}^{n-1} v_j$. If $\sum_{j=1}^{n-1} \sqrt{1-\frac{v_j}{v_n}}$ $\leq$ $n-2$,  then there is a unique $\mu \in (0,  v_n^{-1}]$ solving $h_1(\mu)=n-2$ and the solution for the optimization problem~(\ref{2koptimalproblem}) is
$$p_i = \frac{1+\sqrt{1-\mu v_i}}{2(n-1)}, \>\>\> i=1,\ldots, n.$$
Otherwise, $\sum_{j=1}^{n-1} \sqrt{1-\frac{v_j}{v_n}} > n-2$,  then there is a unique $\mu \in (0,  v_n^{-1})$ solving $h_2(\mu)=n-2$ and the solution for the problem~(\ref{2koptimalproblem}) is
$$p_i = \frac{1+\sqrt{1-\mu v_i}}{2(n-1)}, \>\>\> i=1,\ldots, n-1;\>\>\> p_n =  \frac{1-\sqrt{1-\mu v_n}}{2(n-1)}\ .$$
For both cases, $f$ attains its maximum
$$f(p_1, \ldots, p_n) = p_1\cdots p_n \left[\frac{v_i}{p_i} + \frac{4(n-1)^2 p_i}{\mu}\right], \>\>\> i=1,\ldots, n.$$
\end{theorem}

\begin{example}\label{2^3example}{\rm
Let $n=8$ and $v_j=j$, $j=1, \ldots, 8$. Then $0<v_1< \cdots < v_8 < v_1 + \cdots + v_8$ and $\sum_{j=1}^{n-1}\sqrt{1-\frac{v_j}{v_n}} \leq n-2$ are satisfied. The numerical solution of $h_1(\mu)=n-2$ is
$\mu=0.09260780864$. Based on Theorem~\ref{2ktheorem}, $p_1 =0.1394693827$, $p_2 =0.1359038626$, $p_3 =0.1321292663$, $p_4 =0.1281038353$, $p_5 =0.1237697284$, $p_6 =0.1190427279$, $p_7 =0.1137915161$, $p_8 =0.1077896806$, and the maximum of $f$ is $0.00001753019048$.
}\end{example}

\begin{remark}\label{2kremark}{\rm
Theorem~\ref{2ktheorem} provides an alternative approach for the optimization problem~(\ref{22optimalproblem}), although the answer provided here is not totally analytic ($\mu$ needs to be found numerically by solving an equation of $\mu$, either $h_1(\mu)=n-2$ or $h_2(\mu)=n-2$).
}\end{remark}

Now we remove the assumption that $v_i>0$ for all $i$. Since $v_i = |X[1, \ldots,$ $i-1,$ $i+1,\ldots, n]|^2 w_1\cdots w_{i-1}w_{i+1}\cdots w_n$, this assumption is true only if no row of $X$ can be written as a linear combination of $(n-2)$ other rows. Otherwise, there might be a row of $X$ which is a linear combination of $s$ other rows, where $1\leq s\leq n-2$. For typical applications, the first column of the design matrix $X$ is a vector of $1$'s. In that case, $s=1$ violates that the rows of $X$ are distinct. So we allow $2\leq s\leq n-2$. Without any loss of generality, we may assume the $(n-s)$th row of $X$ is a linear combination of the rows below it. The lemma as follows asserts that $v_1=\cdots=v_{n-s-1}=0$.

\begin{lemma}\label{2kgeneralv}
Let ${\bf x}_1, \ldots, {\bf x}_n$ denote the rows of $X$. Assume that ${\bf x}_i$'s are distinct and rank$(X)=n-1$. Suppose ${\bf x}_{l+1} = c_{l+2} {\bf x}_{l+2} + \cdots + c_n {\bf x}_n$, where $1\leq l\leq n-3$, $c_i \neq 0, i=l+2, \ldots, n$. Then $v_1=\cdots = v_l=0$ and $v_i>0$ for $i=l+1, \ldots, n$.
\end{lemma}

Given that $0=v_1=\cdots=v_l<v_{l+1} \leq \cdots \leq v_n$, the same arguments towards Theorem~\ref{2ktheorem} till equations~(\ref{2kmu0}) are still valid. Based on (\ref{2kmu0}), we immediately obtain $p_i = \frac{1}{n-1}$ for $i=1, \ldots, l$. Then equations~(\ref{2kmu}) and the arguments afterwards are still valid if we restrict statements on $i=l+1, \ldots, n$ only. Thus a theorem similar to Theorem~\ref{2ktheorem} while dealing with degenerated ${\bf x}_i$'s is obtained as follows.

\begin{theorem}\label{2ktheoremgeneral}
Assume that $0 = v_1 = \cdots = v_l < v_{l+1} \leq \cdots \leq v_n < \sum_{j=1}^{n-1} v_j$, where $1\leq l\leq n-3$. If $\sum_{j=1}^{n-1} \sqrt{1-\frac{v_j}{v_n}}\leq n-2$,  then there is a unique $\mu \in (0,  v_n^{-1}]$ solving $h_1(\mu)=n-2$ and the solution for the optimization problem~(\ref{2koptimalproblem}) is
$$p_1=\cdots=p_l=\frac{1}{n-1};\>\> p_i = \frac{1+\sqrt{1-\mu v_i}}{2(n-1)}, \>\>\> i=l+1,\ldots, n.$$
Otherwise, $\sum_{j=1}^{n-1} \sqrt{1-\frac{v_j}{v_n}} > n-2$,  then there is a unique $\mu \in (0,  v_n^{-1})$ solving $h_2(\mu)=n-2$ and the solution for the problem~(\ref{2koptimalproblem}) is $p_1=\cdots=p_l=1/(n-1)$;
$$p_i = \frac{1+\sqrt{1-\mu v_i}}{2(n-1)}, \>\>\> i=l+1,\ldots, n-1;\>\> p_n =  \frac{1-\sqrt{1-\mu v_n}}{2(n-1)}.$$
For both cases, $f$ attains its maximum $4(n-1)p_1\cdots p_n/\mu$.
\end{theorem}

\section{Bridging the Gap between Continuous and Discrete Factors}
\label{section_continuous}

In this section, we aim to make connections between D-optimal designs with quantitative factors and D-optimal designs with pre-specified set of design points, to which our results in previous sections can be applied.

Again, we consider an experiment with response $Y$ from a single-parameter exponential family and two factors labeled by $x_1, x_2$ respectively. Suppose $Y$ is modeled by a generalized linear model with link function $g$, that is,
$g(E(Y)) = \eta = \beta_0 + \beta_1x_1 + \beta_2 x_2$~.

In this section, we assume that the two factors $x_1$ and $x_2$ are quantitative or continuous, $x_1 \in [a_1, b_1]$ and $x_2 \in [a_2, b_2]$. Following Stufken and Yang (2012), the D-optimal design problem here is to find the optimal set of design points $(x_{i1}, x_{i2}) \in [a_1, b_1]\times [a_2, b_2]$, $i=1, \ldots, m$, along with the corresponding allocation $(p_1, \ldots, p_m)'$, where $m \geq 3$ is not fixed. The objective function still takes the form of $|X'WX|$ with $X=({\bf x}_1, {\bf x}_2, \ldots, {\bf x}_m)'$ and $W={\rm Diag}\{p_1 w_1, \ldots, p_m w_m\}$, where ${\bf x}_i=(1, x_{i1}, x_{i2})'$, $i=1, \ldots, m$. Note that $w_i = \nu({\bf x}_i'\boldsymbol{\beta})$, $i=1, \ldots, m$, where $\nu = \left[(g^{-1})'\right]^2/r$ with $r(\eta)={\rm Var}(Y)$ (see Yang and Mandal (2013) for more details), and $\boldsymbol{\beta}=(\beta_0, \beta_1, \beta_2)'$ is assumed to be known for locally optimal design problems.

\begin{lemma}\label{lm:ab11}
The D-optimal design problem with $x_1 \in [a_1, b_1], x_2 \in [a_2, b_2]$ and parameters $\beta_0, \beta_1, \beta_2$ is equivalent to the D-optimal design problem with $x_1^* \in [-1, 1], x_2^* \in [-1, 1]$ and parameters $\beta_0^*, \beta_1^*, \beta_2^*$, where $x_1^* = (2x_1 - a_1 - b_1)/(b_1-a_1)$, $x_2^* = (2 x_2 - a_2 - b_2)/(b_2 - a_2)$, $\beta_0^* = \beta_0 + \beta_1(a_1+b_1)/2 + \beta_2(a_2+b_2)/2$, $\beta_1^* = \beta_1(b_1 - a_1)/2$, $\beta_2^* = \beta_2(b_2-a_2)/2$.
\end{lemma}

According to Lemma~\ref{lm:ab11}, in order to solve the original design problem with $x_1 \in [a_1, b_1], x_2 \in [a_2, b_2]$ and parameters $\beta_0, \beta_1, \beta_2$, one can always do linear transformations and solve the corresponding design problem with $x_1^* \in [-1, 1], x_2^* \in [-1, 1]$ and parameters $\beta_0^*, \beta_1^*, \beta_2^*$. If one obtains a D-optimal design $\{((x_{i1}^*, x_{i2}^*), p_i)\}_{i=1, \ldots, m}$ for the transformed design problem, then $\{((x_{i1}, x_{i2}),$ $p_i)\}_{i=1, \ldots, m}$ is a D-optimal design for the original problem, where $x_{i1} = (a_1+b_1)/2 + x_{i1}^*(b_1 - a_1)/2$, $x_{i2} = (a_2 + b_2)/2 + x_{i2}^*(b_2 - a_2)/2$.

\bigskip
From now on, we assume $a_1=a_2=-1$ and $b_1=b_2=1$ to simplify the notations. An interesting design question with two quantitative factors $x_1, x_2 \in [-1, 1]$ is when the set of boundary points $\{(1, 1), (1, -1), (-1, 1), (-1, -1)\}$ is a D-optimal set of design points. In that case, the experimenter only needs to consider the boundary points during the experiment.

\begin{theorem}\label{thm:quantitative}
Consider a design problem under a generalized linear model with two quantitative factors with levels $x_1, x_2 \in [-1, 1]$. The D-optimal design can be constructed on the four boundary points only, that is, $\xi=\{((1,1),p_1), ((1,-1),$ $p_2),$ $((-1,1),p_3), ((-1,-1), p_4)\}$ is a D-optimal design for some allocation $(p_1,$ $p_2,$ $p_3, p_4)$, if and only if $(p_1, p_2, p_3, p_4, 0)$ is a D-optimal allocation for the design problem with pre-specified design matrix
\begin{equation}\label{X5mat}
X=\left(
\begin{array}{rrr}
1 &  1 &  1\\
1 &  1 & -1\\
1 & -1 &  1\\
1 & -1 & -1\\
1 &  a &  b
\end{array}\right)
\end{equation}
for any $a,b\in [-1,1]$.
\end{theorem}

The proof of Theorem~\ref{thm:quantitative} is arranged in Appendix. Now we derive a more explicit condition of Theorem~\ref{thm:quantitative} which is easier to be justified in practice. Based on Yang, Mandal and Majumdar (2013, Lemma~3.1), the objective function of the design with design matrix $X$ defined as in (\ref{X5mat}) is
\begin{eqnarray*}
|X'WX| &=& 16 q_1 q_2 q_3 + 16 q_1 q_2 q_4 + 16 q_1 q_3 q_4 + 16 q_2 q_3 q_4\\
 &+& 4(1 - a)^2 q_1 q_2 q_5 + 4(1 - b)^2 q_1 q_3 q_5 + 4(a + b)^2 q_2 q_3 q_5\\
 &+& 4(a - b)^2 q_1 q_4 q_5 + 4(1 + b)^2 q_2 q_4 q_5 + 4(1 + a)^2 q_3 q_4 q_5
\end{eqnarray*}
where $q_i=p_iw_i$, $i=1,2,\ldots, 5$.

Let $\mathbf{p}_{50}=(p_1, p_2, p_3, p_4, 0)'$, that is, a design restricted to the four boundary points. Then $f(\mathbf{p}_{50})= |X'WX|= 16(p_1 p_2 p_3 w_1 w_2 w_3 + p_1 p_2 p_4 w_1 w_2 w_4 + p_1 p_3 p_4 w_1 w_3 w_4 +  p_2 p_3 p_4 w_2 w_3 w_4)$. Following Yang, Mandal and Majumdar (2013), we define for $i=1,2,\ldots, 5$ and $0\leq z\leq 1$,
\begin{equation}\label{fiz}
f_i(z) = f\left(\frac{1-z}{1-p_i}p_1, \ldots, \frac{1-z}{1-p_i}p_{i-1}, z, \frac{1-z}{1-p_i}p_{i+1}, \ldots, \frac{1-z}{1-p_i}p_5\right).
\end{equation}
Applying Theorem~3.1 in Yang and Mandal (2013) to our case, we need to check whether or not $f_5(1/2) \leq f({\mathbf p}_{50})/2$. It can be verified that $f(\mathbf{p}_{50}) - 2 f_5(\frac{1}{2}) = 3f(\mathbf{p}_{50})/4 - w_5(a,b) h(a,b)$, where
\begin{eqnarray}
h(a,b)
&=& p_1 p_2 w_1 w_2 + p_1 p_3 w_1 w_3 + p_2 p_4 w_2 w_4 + p_3 p_4 w_3 w_4\nonumber\\
& & + b^2 (p_1 p_3 w_1 w_3 + p_2 p_3 w_2 w_3 + p_1 p_4 w_1 w_4 + p_2 p_4 w_2 w_4)\nonumber\\
& & + 2 b (- p_1 p_3 w_1 w_3 +  p_2 p_4 w_2 w_4)\nonumber\\
& & + a^2 (p_1 p_2 w_1 w_2 + p_2 p_3 w_2 w_3 + p_1 p_4 w_1 w_4 + p_3 p_4 w_3 w_4)\nonumber\\
& & + 2 a (- p_1 p_2 w_1 w_2 + p_3 p_4 w_3 w_4)\nonumber\\
& & + 2 a b (p_2 p_3 w_2 w_3 - p_1 p_4 w_1 w_4)\label{h(a,b)}
\end{eqnarray}
Note that $w_5=w_5(a,b)=\nu(\beta_0+a\beta_1+b\beta_2)$ is a function of $a,b$, while
$w_1 = \nu(\beta_0+\beta_1+\beta_2)$,
$w_2 = \nu(\beta_0+\beta_1-\beta_2)$,
$w_3 = \nu(\beta_0-\beta_1+\beta_2)$,
$w_4 = \nu(\beta_0-\beta_1-\beta_2)$ do not depend on $a,b$.
With the aid of $h(a,b)$, we are able to express the condition of Theorem~\ref{thm:quantitative} in a more explicit way. The preceding arguments prove the following theorem.

\begin{theorem}\label{thm:22boundary}
Given $\beta_0, \beta_1, \beta_2$, a D-optimal design with quantitative factors $x_1, x_2$ $\in$ $[-1, 1]$ could be constructed only on the four boundary design points $\{(1,1), (1,-1),$ $(-1,1), (-1,-1)\}$ if and only if
\begin{equation}\label{boundarypointscondition}
\nu(\beta_0+a\beta_1+b\beta_2) h(a,b) \leq \frac{3}{4}f({\mathbf p}_4)\ ,\mbox{ for all }a,b\in [-1,\ 1],
\end{equation}
where $h(a,b)$ is defined as in (\ref{h(a,b)}), $\mathbf{p}_4=(p_1,p_2,p_3,p_4)'$ is the locally D-optimal allocation for the $2^2$ main-effects model, and $f(\mathbf{p}_4) = 16(p_1 p_2 p_3 w_1 w_2 w_3$ $+$ $p_1 p_2 p_4 w_1 w_2 w_4$ $+$ $p_1 p_3 p_4 w_1 w_3 w_4 +  p_2 p_3 p_4 w_2 w_3 w_4)$ is the value of the $2^2$ main-effects design problem.
\end{theorem}

Note that ${\mathbf p}_4$ and $f({\mathbf p}_4)$ in Theorem~\ref{thm:22boundary} can be calculated analytically according to Theorem~\ref{thm:22solution}. Then the inequality~(\ref{boundarypointscondition}) is a known function of $a$ and $b$ only. Numerical approaches could be used for checking if the inequality is valid or not. The analytic solution derived in Section~\ref{section_22} turns out to be critical for applying Theorem~\ref{thm:22boundary} (see Section~\ref{section_simu}.2).

\section{Applications of Analytic Solutions}
\label{section_simu}

\subsection{Significance of analytic solutions}

We first show that our analytic approaches reduce computational time significantly. Three types of ``optimal'' allocations are under comparison: (i) analytic ones, ${\mathbf p}_a$ for two factors based on Theorem~\ref{thm:22solution} or ${\mathbf p}_e$ for $k$ factors ($k\geq 3$) based on Theorem~\ref{2ktheorem}; (ii) ${\mathbf p}_s$ based on a quasi-Newton method used by Yang, Mandal and Majumdar (2012); (iii) ${\mathbf p}_l$ based on the lift-one algorithm proposed by Yang, Mandal and Majumdar (2013) which works much faster and more accurate than commonly used nonlinear optimization algorithms.

Table~\ref{table22computation} lists the computational times of ${\mathbf p}_a, {\mathbf p}_s, {\mathbf p}_l$ for 10,000 cases with $\beta_i$'s simulated i.i.d.~from uniform or normal distribution under $2^2$ main-effects model with logit link. The analytic ${\mathbf p}_a$ run significantly faster than the numerical ones. The difference tends to be larger as the variance of the distribution increases. It is because the proportion of extreme $\beta_i$'s become larger which leads to more saturated cases (see Yang, Mandal and Majumdar (2012)). The searching time needed by typical nonlinear numerical algorithms such as quasi-Newton is much longer for a solution at the boundary. The lift-one algorithm is not affected much by the saturated cases. Table~\ref{table2kcomputation} shows the change of computational times along with the number $k$ of factors. As for $k=6$, the original life-one algorithm suffers numerical errors due to the large number of parameters, while our analytic approach is not affected much. All the computational time costs here are recorded on a Windows~7 PC with Intel(R) Core(TM) i5-2400 CPU at 3.10GHz and 4GB memory.

\begin{table}[ht]
\caption{Time cost in secs for 10,000 simulations for $2^2$ design}
\label{table22computation}
\begin{center}
\begin{tabular}{c|r|r|r|r|r|r}\hline
Solution        & U$(-1,1)$  & U$(-2,2)$ & U$(-3,3)$ & $N(0,1)$ & $N(0,2)$ & $N(0,3)$\\ \hline
${\mathbf p}_a$ & 3.81          & 2.15         & 1.35         & 2.90     & 1.59     & 1.05\\
${\mathbf p}_s$ & 9.54          & 17.40        & 20.78        & 13.78    & 20.37    & 21.86\\
${\mathbf p}_l$ & 9.43          & 10.54        & 10.85        & 10.18    & 11.00    & 11.04\\ \hline
\end{tabular}
\end{center}
\end{table}

\begin{table}[ht]
\caption{Time cost in secs for 10,000 simulations ($U(-3,3)$) for $2^k$ design}
\label{table2kcomputation}
\begin{center}
\begin{tabular}{c|r|r|r|r|r}\hline
$k$             & 2             & 3            & 4            & 5        & 6   \\ \hline
${\mathbf p}_e$ & 1.40          & 1.31         & 1.34         & 1.75     & 4.11\\
${\mathbf p}_l$ & 10.85         & 23.51        & 50.14        & 135.41   & $-$ \\ \hline
\end{tabular}
\end{center}
\end{table}

\begin{figure}[ht]
\begin{center}
\psfig{figure=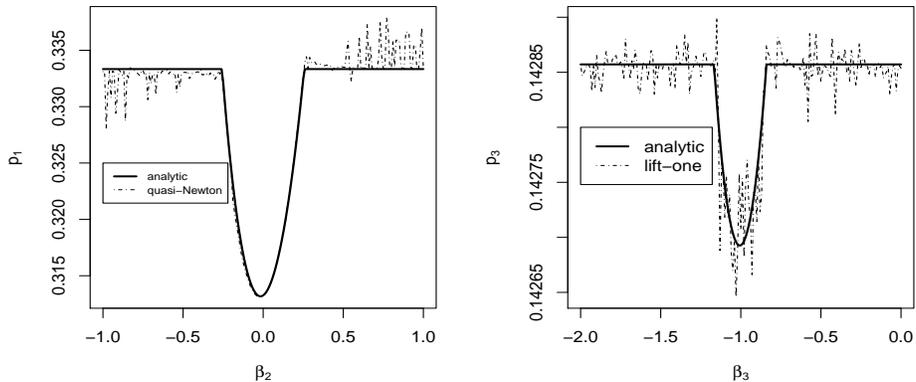,height=2.5in,width=5in,angle=0}
\vskip -.5cm
\captionsetup{width=.75\textwidth}
\caption{Comparison of allocation solutions with different $\beta$'s: optimal $p_1$ under the $2^2$ model: $\eta = -2 + x_1 + \beta_2 x_2$ with $\beta_2 \in [-1,1]$ (left panel); optimal $p_3$ under the $2^3$ model: $\eta = -3 + x_1 -2 x_2 + \beta_3 x_3 - x_1x_2 + x_1 x_3 + 2 x_2 x_3$ with $\beta_3 \in [-2,0]$ (right panel)}\label{palong}
\end{center}
\end{figure}

Secondly, we show the advantage of the analytic approaches over the numerical ones in terms of accuracy. Although numerical solutions can be highly efficient since the value of the objective function $f({\mathbf p})$ is typically the target of the algorithm, the behavior of numerically optimal allocations may not be satisfying at all. Figure~\ref{palong} shows the comparison of allocations in terms of changes along with parameter values. The numerical solutions (quasi-Newton or lift-one) may wiggly around the analytic one as $\beta_i$ changes, even they are highly efficient ($f({\mathbf p}_s)/f({\mathbf p}_a)$, $f({\mathbf p}_l)/f({\mathbf p}_e) > 99.99\%$). They may be misleading when one wants to study how the optimal allocation changes along with parameters. It is critical for locally optimal designs with assumed values of parameters.

\subsection{Identify region of parameters for boundary designs}

\begin{figure}[ht]
\begin{center}
\psfig{figure=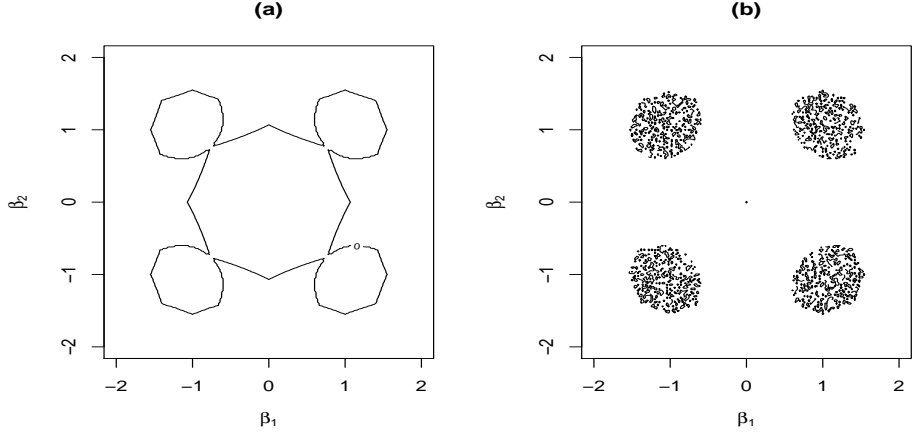,height=2.5in,width=5in,angle=0}
\vskip -.5cm
\captionsetup{width=.75\textwidth}
\caption{Region of $(\beta_1,\beta_2)$ ($\beta_0=-1$) such that a D-optimal design with factors $x_1,x_2\in [-1,1]$ could be constructed on boundary points only: (a) Based on optimal allocation ${\mathbf p}_a$; (b) Based on allocation ${\mathbf p}_l$}\label{figthm3a}
\end{center}
\end{figure}

\begin{figure}[ht]
\begin{center}
\psfig{figure=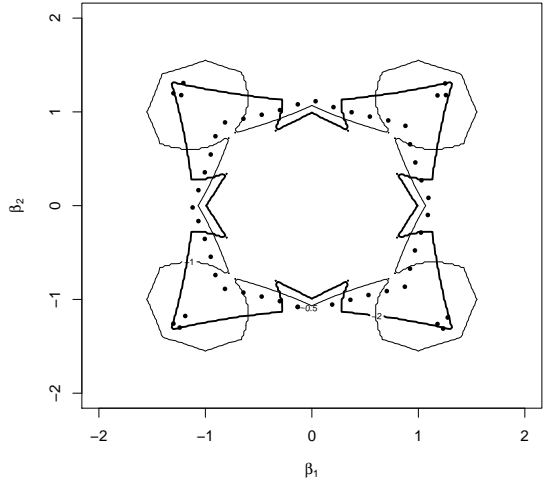,height=3in,width=3in,angle=0}
\vskip -.3cm
\captionsetup{width=.75\textwidth}
\caption{Region of $(\beta_1,\beta_2)$ such that a D-optimal design with factors $x_1,x_2\in [-1,1]$ could be constructed on boundary points only: Solid line ($\beta_0=1$), thick solid line ($\beta_0=-2$), and dot line ($\beta_0=-0.5)$ which consists of 5 disjoint pieces.}\label{figthm3b}
\end{center}
\end{figure}

Although the numerical allocations can be highly efficient with respect to the analytical ones, the tiny difference matters when highly precise solution is needed. For example, in order to apply Theorem~\ref{thm:22boundary}, one needs to check if (\ref{boundarypointscondition}) is true for all $a,b\in [-1,1]$. Let
$$s(a,b) = \frac{3}{4} f({\mathbf p}) - \nu(\beta_0 + a \beta_1 + b \beta_2) h(a,b),\>\> a,b \in [-1,1].$$
Since $s(a,b)$ is differentiable for typical link functions, nonlinear optimization such as quasi-Newton method with box constrains (Byrd, Lu, Nocedal and Zhu, 1995) works well in finding the minimum of $s(a,b)$. If $\min s = 0$, then a D-optimal design could be constructed on boundary points only. The critical part is to calculate optimal ${\mathbf p}$ and $f({\mathbf p})$ precisely. To illustrate the significance of ${\mathbf p}_a$ or ${\mathbf p}_e$, we fix $\beta_0=-1$ and vary $\beta_1, \beta_2$ from $-2$ to $2$. For each combination $(\beta_0, \beta_1, \beta_2)$, we use either ${\mathbf p}_a$ or ${\mathbf p}_l$ for $s(a,b)$ before its minimization. One can see from Figure~\ref{figthm3a} that a reasonable region of $(\beta_1, \beta_2)$ is built up based on ${\mathbf p}_a$ (see Figure~\ref{figthm3a}(a)) while a failure occurs with the use of ${\mathbf p}_l$ (see Figure~\ref{figthm3a}(b)). For boundary lines of the regions with other values of $\beta_0$, please see Figure~\ref{figthm3b}.

\appendix

\section{Appendix}

\subsection{Proof of Lemma~\ref{y1>1}}

Let $h(y_1)=c_0 + c_1 y_1 + c_2 y_1^2 + c_3 y_1^3 + c_4 y_1^4$. Note that $h(-\infty) = \infty$, $h\left(-v_1/v_4\right) = -4v_1^3(v_2-v_3)^2/v_4 < 0$, $h(0) = c_0 > 0$, $h(1) = -(v_1-v_4)^2[(v_1+v_4)^2 -$ $(v_2-v_3)^2]$ $< 0$, $h(\infty) = \infty$. Therefore, $h(y_1)=0$ has four real roots in $(-\infty, -v_1/v_4)$, $(-v_1/v_4, 0)$, $(0, 1)$, and $(1,\infty)$, respectively. The only solution $y_1>1$ is what we need to solve (\ref{22optimalproblem}).
\hfill{$\#$}

\subsection{Proof of Lemma~\ref{2klemma}}

Since $p_n = 1-\sum_{i=1}^{n-1} p_i$, then
\begin{eqnarray*}
& & f(p_1, \ldots, p_n)\\
&=&\sum_{j=1}^{n-1}p_1p_2\cdots p_{n-1}\left(1-\sum_{i=1}^{n-1}p_i\right)\frac{v_j}{p_j} + p_1p_2\ldots p_{n-1}v_n\\
&=& \sum_{j=1}^{n-1}p_1\cdots p_{j-1}p_{j+1}\cdots p_{n-1}\left(1-\sum_{i=1}^{n-1}p_i+p_j\right)v_j + p_1p_2\ldots p_{n-1}\left(v_n-\sum_{j=1}^{n-1}v_j\right)\\
&\leq & \sum_{j=1}^{n-1} \left(\frac{1}{n-1}\right)^{n-1}v_j + \left(\frac{p_1 + p_2 + \cdots + p_{n-1}}{n-1}\right)^{n-1} \left(v_n-\sum_{j=1}^{n-1}v_j\right)\\
&\leq & \frac{1}{(n-1)^{n-1}}\sum_{j=1}^{n-1} v_j + \frac{1}{(n-1)^{n-1}} \left(v_n-\sum_{j=1}^{n-1}v_j\right)\\
&=& \frac{v_n}{(n-1)^{n-1}}\\
&=& f(\frac{1}{n-1}, \ldots, \frac{1}{n-1}, 0)\ .
\end{eqnarray*}
Based on the inequality of arithmetic and geometric means, the two ``$\leq$" above are both ``$=$" if and only if $p_1 = p_2 = \cdots = p_{n-1} = \frac{1}{n-1}, p_n = 0$.
\hfill{$\#$}

\subsection{Proof of Lemma~\ref{2klemmab}}

If $p_i=0$, then $f(\mathbf{p})=v_i p_1\cdots p_{i-1}p_{i+1}\cdots p_n$ which attains its maximum $\frac{v_i}{(n-1)^{n-1}}$ at $p_1 = \cdots = p_{i-1} = p_{i+1} = p_n=\frac{1}{n-1}, p_i=0$. The largest value across different $i$'s is $\frac{v_n}{(n-1)^{n-1}}$ at $i=n$. On the other hand, set
$$F(t) = f\left(\frac{1-t}{n-1}, \ldots, \frac{1-t}{n-1}, t\right) = t\left(\frac{1-t}{n-1}\right)^{n-2} \sum_{j=1}^{n-1} v_j + \left(\frac{1-t}{n-1}\right)^{n-1} v_n\ .$$
Note that $F'(0) = \frac{1}{(n-1)^{n-2}}(\sum_{j=1}^{n-1} v_j - v_n) > 0$. $F(t)$ won't attain its maximum at $t=0$ which implies that $f({\mathbf p})$ won't attains its maximum at $(\frac{1}{n-1}, \ldots, \frac{1}{n-1}, 0)'$. Therefore, $f(\mathbf{p})$ won't attain its maximum at any boundary point.
\hfill{$\#$}

\subsection{Proof of Theorem~\ref{2ktheorem}}

We only need to show the existence and uniqueness of $\mu$.

If $\sum_{j=1}^{n-1}\sqrt{1-\frac{v_j}{v_n}} \leq n-2$, then $h_1(v_n^{-1}) \leq n-2$. Since $h_1(0)=n$ and $h_1$ is a strictly decreasing continuous function, there exists a unique solution of $h_1(\mu) = n-2$ with $\mu \in (0, v_n^{-1}]$.

If $\sum_{j=1}^{n-1}\sqrt{1-\frac{v_j}{v_n}} > n-2$, then $h_2(v_n^{-1}) > n-2$. Since $h_2(0)=n-2$, $h'_2(0)$ $=$ $\frac{1}{2}(v_n$ $- \sum_{j=1}^{n-1} v_j) < 0$, and $h_2$ is continuous, then $h_2(\mu) = n-2$ admits a solution in $(0, v_n^{-1})$. In order to show that the solution is unique, let
$$g_2(\mu) = 2\sqrt{1-\mu v_n} h'_2(\mu) = v_n - \sum_{j=1}^{n-1} v_j\sqrt{\frac{1-\mu v_n}{1-\mu v_j}}\ .$$
Then $g'_2(\mu)=\frac{1}{2}\sum_{j=1}^{n-1} v_j \sqrt{\frac{1-\mu v_j}{1-\mu v_n}} \cdot \frac{v_n - v_j}{(1-\mu v_j)^2} > 0$ for $\mu \in (0, v_n^{-1})$. Since $g_2(0)=2h'_2(0) < 0$ and $g_2(v_n^{-1}) = v_n > 0$, then $g_2(\mu)=0$ for one and only one $\mu \in (0, v_n^{-1})$. Therefore $h'_2(\mu)=0$ for one and only one $\mu\in (0, v_n^{-1})$ which is for a local minimum of $h_2$. The conclusion is that $h_2(\mu)=n-2$ only admits one positive solution in $(0, v_n^{-1})$.

Since $\lambda=\partial f/\partial p_i$, $i=1, \ldots, n$, then $f(p_1, \ldots, p_n) = \lambda p_i + p_1\cdots p_n v_i/p_i$ which could be used conveniently for calculating $f(p_1, \ldots, p_n)$.
\hfill{$\#$}

\subsection{Proof of Lemma~\ref{2kgeneralv}}

Since ${\bf x}_{l+1} = c_{l+2} {\bf x}_{l+2} + \cdots + c_n {\bf x}_n$ and rank$(X)=n-1$, then ${\bf x}_1, \ldots, {\bf x}_l$, ${\bf x}_{l+2}$, $\ldots$, ${\bf x}_n$ are linearly independent, which implies $|X[1, \ldots, l, l+2, \ldots, n]|$ $\neq$ $0$ and $v_{l+1} >0$. For $i=1,\ldots, l$, $|X[1,\ldots, i-1, i+1, \ldots, n]| = 0$ and thus $v_i = 0$ due to ${\bf x}_{l+1} = c_{l+2} {\bf x}_{l+2} + \cdots + c_n {\bf x}_n$.

For $i=l+2, \ldots, n$, since ${\bf x}_1, \ldots, {\bf x}_l, {\bf x}_{l+2}, \ldots, {\bf x}_{i-1}, {\bf x}_{i+1}, \ldots {\bf x}_n$ are linearly independent too. If $|X[1, \ldots, l+1, \ldots, i-1, i+1, \ldots, n]|=0$, then \begin{equation}\label{x(l+1)2}
{\bf x}_{l+1} = c'_1{\bf x}_1 + \cdots + c'_l {\bf x}_l + c'_{l+2} {\bf x}_{l+2} + \cdots + c'_{i-1} {\bf x}_{l-1} + c'_{i+1} {\bf x}_{i+1} + \cdots + c'_n {\bf x}_n
\end{equation}
for some $c'_1, \ldots, c'_l, c'_{l+2}, \ldots, c'_{i-1}, c'_{i+1}, \ldots, c'_n$. Due to linear independence of ${\bf x}_1, \ldots, {\bf x}_l,$ ${\bf x}_{l+2},$ $\ldots,$ ${\bf x}_n$, $c_{l+2} {\bf x}_{l+2} + \cdots + c_n {\bf x}_n$ should be the unique linear expression of ${\bf x}_{l+1}$. It implies expression~(\ref{x(l+1)2}) is not possible which does not include ${\bf x}_i$. The contradiction leads to $|X[1, \ldots, l+1, \ldots, i-1, i+1, \ldots, n]|\neq 0$. That is, $v_i > 0$ for $i=l+2, \ldots, n$.
\hfill{$\#$}

\subsection{Proof of Lemma~\ref{lm:ab11}}

Let $\{({\bf x}_i, p_i)\}_{i=1,\ldots, m}$ be an arbitrary design for the original design problem, where ${\bf x}_i = (1, x_{i1}, x_{i2})'$, $i=1, \ldots, m$. Define ${\bf x}^*_i = (1, x_{i1}^*, x_{i2}^*)'$, $i=1, \ldots, m$ be the transformed supporting points, that is, $x_{i1}^* = \frac{2x_{i1} - a_1 - b_1}{b_1-a_1} \in [-1, 1]$, $x_{i2}^* = \frac{2x_{i2} - a_2 - b_2}{b_2 - a_2} \in [-1, 1]$. It can be verified that $\eta_i^* = \beta_0^* + \beta_1^* x_{i1}^* + \beta_2^* x_{i2}^* = \beta_0 + \beta_1 x_1 + \beta_2 x_2 = \eta_i$, $i=1, \ldots, m$. Then $w_i^* = \nu(\eta^*_i)=\nu(\eta_i) = w_i$ and $W_*= {\rm Diag}\{p_1 w_1^*, \ldots, p_m w_m^*\} = {\rm Diag}\{p_1 w_1, \ldots, p_m w_m\}$ for the same set of allocations $p_1, \ldots, p_m$.

On the other hand, the transformed design matrix $X_* = ({\bf x}_1^*, \ldots, {\bf x}_m^*)' = XT$, where the transformation matrix
\[
T=\left(\begin{array}{ccc}
1 & -\frac{b_1+a_1}{b_1-a_1} & -\frac{b_2 + a_2}{b_2-a_2}\\
0 & \frac{2}{b_1 - a_1} & 0\\
0 & 0 & \frac{2}{b_2-a_2}
\end{array}\right).
\]
The transformed design problem is to maximize $|X_*'W_*X_*|=|T'X'WXT|$ $=$ $|T|^2\cdot |X'WX|$, where $|T| = \frac{4}{(b_1 - a_1)(b_2 - a_2)}$ is a constant. Thus the transformed D-optimal design problem is equivalent to the original D-optimal design problem. Actually, $\{({\bf x}_i, p_i)\}_{i=1,\ldots, m}$ is D-optimal for the original problem if and only if $\{({\bf x}_i^*, p_i)\}_{i=1,\ldots, m}$ is D-optimal for the transformed design problem.
\hfill{$\#$}

\subsection{Proof of Theorem~\ref{thm:quantitative}}

The ``only if'' part is straightforward. For the ``if" part, let $\{ {\bf x}_1, \ldots, {\bf x}_m\}$ be the set of supporting points of a D-optimal design with maximum determinant $d_m=|X_m'W_mX_m| = |\sum_{i=1}^m q_i \nu({\bf x}_i'\boldsymbol{\beta}) {\bf x}_i {\bf x}_i'|$, where $q_1, \ldots, q_m$ are the corresponding D-optimal allocations. Let ${\bf z}_1=(1, 1, 1)', {\bf z}_2 = (1, 1, -1)', {\bf z}_3=(1, -1, 1)', {\bf z}_4 = (1, -1, -1)'$. Combine ${\bf z}_1, \ldots, {\bf z}_4$ and ${\bf x}_1, \ldots,$ ${\bf x}_m$ into ${\bf z_1}$, $\ldots$, ${\bf z}_l$ after removing duplicated supporting points. Then $\max\{m, 4\} \leq l\leq m+4$. Suppose $(p_1^*, \ldots, p_l^*)'$ is a D-optimal allocation of the design problem with pre-specified supporting points ${\bf z}_1, \ldots, {\bf z}_l$, then the maximum determinant is $d_l = |X_l' W_l X_l| = |\sum_{i=1}^l p_i^* \nu({\bf z}_i'\boldsymbol{\beta}) {\bf z}_i {\bf z}_i'|$. Since ${\bf x}_1, \ldots, {\bf x}_m$ are part of ${\bf z}_1, \ldots, {\bf z}_l$, then $d_m=d_l$.

To show that the D-optimal design with two quantitative factors can be constructed only on the boundary points with optimal allocations $p_1, \ldots, p_4$, we only need to show that $(p_1, \ldots, p_4, 0, \ldots, 0)'$ achieves $d_l$ for the design problem with pre-specified supporting points ${\bf z}_1, \ldots, {\bf z}_l$. Applying Theorem~3.1 of Yang and Mandal (2013), then $$f({\mathbf p})= \left|\sum_{i=1}^4 p_i \nu({\bf z}_i'\boldsymbol{\beta}){\bf z}_i {\bf z}_i'\right| > 0$$ which is equal to the determinant in the design problem with pre-specified supporting points ${\bf z}_1, \ldots, {\bf z}_4$ only. For $i=1, \ldots, 4$, $0<p_i\leq 1/3$ and $f_i(0) = \frac{1-3p_i}{(1-p_i)^3}f({\mathbf p})$ are satisfied because $(p_1, \ldots, p_4)'$ maximizes the design problem with ${\bf z}_1, \ldots, {\bf z}_4$. Here the definition of $f_i$ can also be found in (\ref{fiz}). For $i=5, \ldots, l$, $p_i=0$ and $f_i(1/2) \leq f({\bf p})/2$ are correct because $(p_1, \ldots, p_4, 0)'$ maximizes the design problem with ${\bf z}_1, \ldots, {\bf z}_4, {\bf z}_i$. According to Theorem~3.1 of Yang and Mandal (2013), $(p_1, \ldots, p_4, 0, \ldots, 0)'$ is a D-optimal allocation for the design problem with ${\bf z}_1, \ldots, {\bf z}_l$ and thus achieves $d_l$.
\hfill{$\#$}

\end{document}